\documentclass[preprint,aps,amsmath,amssymb,nofootinbib,tightenlines%,superscriptaddress
]{revtex4}
\usepackage{amsmath,amssymb}

\usepackage{hyperref}
\usepackage{graphicx}
\usepackage{slashed}

\begin{document}

\vspace{3.0cm}
\preprint{\vbox 
{
\hbox{WSU--HEP--1304} 
%\hbox{hep-ph/yymmddd}
}}

\vspace*{2cm}

\title{Neutrino Democratic Masses, Mixing and Incoherence}

%\date{}                                           % Activate to display a given date or no date

\author{Dmitry Zhuridov}%
\email{dmitry.zhuridov@wayne.edu}%
\email{jouridov@mail.ru}%
\affiliation{Department of Physics and Astronomy, Wayne State University,
Detroit, MI 48201}

\date{\today \\ \vspace{1in}}% It is always \today, today,
             %  but any date may be explicitly specified

\begin{abstract}
It is shown that the well established and confirmed neutrino experimental results, such as atmospheric neutrino oscillations and solar neutrino deficit, can be easily explained for ``democratic" left-handed Majorana neutrinos, taking into account the effect of incoherence. In this model the absolute values of the neutrino masses  can be extracted from the atmospheric neutrino oscillation data, which gives the mass spectrum close to \{0.03 eV, 0.03 eV, 0.06 eV\}.  
The predictions of the considered model and its further testing are discussed. 
\end{abstract}

\maketitle

%%%%%%%%%%%%%%%%%%%%%%%%%%%%%%%%%%%%%%%%%%%%%%%%%%%%
\section{Introduction}

The neutrinos were proposed by W. Pauli in 1930~\cite{Pauli:2000ak} and first detected by C.~%Clyde 
Cowan and F.~% Frederick 
Reines in 1956~\cite{Cowan:1992xc}. However the questions on the type (Dirac or Majorana) and values of their masses still remain actual~\cite{PDG2012}. The most-studied and well established phenomena, which are governed by the neutrino masses, are the solar neutrino deficit~\cite{Davis:1968cp} and the atmospheric neutrino oscillations~\cite{PDG2012}.   In this paper I show that the discussed phenomena can be explained within an economical model with three ``democratic" Majorana neutrinos, taking into account their coherence length. If this theory is confirmed, it will provide the answers on the mentioned questions on the neutrino masses. 

Due to large number of (anti)neutrino flux calculations and Monte Carlo methods in the market (see, e.g., Refs.~[21-31] in Ref.~\cite{Ashie:2005ik} and Refs.~[15-21] in Ref.~\cite{Adamson:2013whj}), and large number of discrepancies among them~\cite{Paukkunen:2013qfx,Zhang:2013ela,Mention:2011rk} we will concentrate on the results, which do not strongly depend on the absolute normalization of the data samples.\footnote{Remind 
the significant difference between the results derived in the 3-Dimensional and the 1-D collinear simulations~\cite{Battistoni:1999at}.}

In the next section the neutrino mass model is introduced, and the neutrino masses and mixing are derived.  The general formula for the neutrino oscillations and its applications to the atmospheric and solar neutrinos in the considered model are given in section~\ref{sec:oscillations}. The predictions of this model for other neutrino experiments, and its further testing are outlined in section~\ref{sec:other_experiments}, which is followed by a conclusion.

\section{Neutrino masses and mixing}
Consider the mass term for three left-handed Majorana neutrinos
\begin{eqnarray}\label{eq:Lm}
	\mathcal{L}_m^\nu	=	-\frac{1}{2}\,	\sum_{\alpha\beta}\bar\nu_{\alpha L}^c M_{\alpha\beta}\nu_{\beta L} + {\rm H.c.},
\end{eqnarray}
where $\alpha,\beta=e,\mu,\tau$ are the flavor indices, $c$ denotes charge conjugation, and
\begin{eqnarray}\label{eq:mass_matrices}
	M	=	 m\left( \begin{array}{ccc}
    0 & 1 & 1 \\ 
    1 & 0 & 1 \\ 
    1 & 1 & 0 \\ 
  \end{array} \right)
\end{eqnarray}
is a ``democratic" mass matrix, which is invariant under the permutation group of three elements $S_3$~\cite{Harari:1978yi,Fritzsch:1995dj,Fukugita:1998vn,Fujii:2002jw,Mohapatra:1998rq,Harrison:2003aw,Nicolaidis:2013hxa}.~\footnote{In general, also the second invariant under $S_3$ matrix $\tilde M=\rm{daig}\{ \tilde m, \tilde m, \tilde m \}$ can be considered, e.g., as small perturbation.} 
The matrix $M$ has the eigenvalues 
\begin{eqnarray}\label{eq:eigenvalues}
	\lambda_1=\lambda_2=-m,	\quad	\lambda_3=2m
\end{eqnarray}
and the corresponding eigenvectors
\begin{eqnarray}\label{eq:eigenvectors}
	N_1^T &=&	\frac{1}{\sqrt{2}} (1,-1,0),	\nonumber\\
	N_2^T &=&	\frac{1}{\sqrt{6}} (1,1,-2),	\\
	N_3^T &=&	\frac{1}{\sqrt{3}} (1,1,1),	\nonumber
\end{eqnarray}
which form the leptonic mixing matrix of tri-bimaximal~\cite{Wolfenstein:1978uw,Fritzsch:1995dj,Harrison:2002er,Harrison:2002kp,Xing:2002sw} type
\begin{eqnarray}\label{eq:U}
	U	=	\left(	  \begin{array}{ccc}
     \frac{1}{\sqrt{2}} & \frac{1}{\sqrt{6}} & \frac{1}{\sqrt{3}} \\ 
    -\frac{1}{\sqrt{2}} & \frac{1}{\sqrt{6}} &  \frac{1}{\sqrt{3}}  \\ 
    0 & -\frac{2}{\sqrt{6}} & \frac{1}{\sqrt{3}} \\ 
  \end{array}	\right).	
\end{eqnarray} 
This matrix can be parametrized as
\begin{eqnarray}\label{eq:Uparametr}
	U	=	R_{12}(\theta_{12}) \times R_{23}(\theta_{23})	=	 
	\left(   \begin{array}{ccc}
    c_{12} & s_{12}c_{23} & s_{12}s_{23} \\ 
    -s_{12} & c_{12}c_{23} & c_{12}s_{23} \\ 
    0 & -s_{23} & c_{23} \\ 
  \end{array}	\right)
\end{eqnarray} 
where $c_{ij}\equiv\cos\theta_{ij}$, $s_{ij}\equiv\sin\theta_{ij}$, $\theta_{12}=45^\circ$, $\theta_{23}=\pi/2-\arctan(1/\sqrt{2})\approx54.7^\circ$, 
and we used nonstandard order of multiplication (compare with Ref.~\cite{Xing:2002sw}) of the two Euler rotation matrixes
\begin{eqnarray}%\label{eq:Uparametr}
	R_{12}(\theta_{12}) = 
	\left( \begin{array}{ccc}
    c_{12} & s_{12} & 0 \\ 
    -s_{12} & c_{12} & 0 \\ 
    0 & 0 & 1 \\ 
  \end{array}	\right),	\qquad
	R_{23}(\theta_{23})	=	
	\left(	  \begin{array}{ccc}
    1 & 0 & 0 \\ 
    0 & c_{23} & s_{23} \\ 
    0 & -s_{23} & c_{23} \\ 
  \end{array}	\right).
\end{eqnarray}

The matrix $U$ diagonalizes the mass matrix $M$ as
\begin{eqnarray}%\label{eq:eigenvalues}
	U^T M U 	=	\rm{diag} (\lambda_1, \lambda_2, \lambda_3),
\end{eqnarray}
and makes the transformation from flavor to mass basis
\begin{eqnarray}
	\nu_{\alpha L}=	\sum_{i=1,2,3} U_{\alpha i}\nu_{i L},
\end{eqnarray}
in which  the mass term in Eq.~\eqref{eq:Lm} has a diagonal form
\begin{eqnarray}\label{eq:LmDiag}
	\mathcal{L}_m^\nu	=	-\frac{1}{2}\,	\sum_i	\lambda_i  \bar\nu_{i L}^c \nu_{i L} + {\rm H.c.}
\end{eqnarray}
Now, using the definition
\begin{eqnarray}%\label{eq:LmDiag}
	\nu_{iL}	\equiv	\frac{1}{2}  (1-\gamma_5)	\psi_i
\end{eqnarray}
with the Majorana spinors $\psi_i=\psi_i^c$, Eq.~\eqref{eq:LmDiag} can be rewritten as
\begin{eqnarray}\label{eq:Lmass}
	\mathcal{L}_m^\nu	=	-\frac{1}{2}\,	\sum_i	\lambda_i  \bar\psi_i \psi_i	=	-\frac{1}{2}\,	\sum_i	s_i m_i  \bar\psi_i \psi_i,
\end{eqnarray}
where $m_1=m_2=m_3/2\equiv m$, and $s_1=s_2=-s_3=-1$ are the sign factors, which can be absorbed by the transformation $\psi_j \to i\gamma_5\psi_j^\prime$ ($j=1,2$). On relation of the sign factors to the neutrino $CP$ properties and mixing matrix see section 2.3.2 in Ref.~\cite{Doi:1985dx} and references therein. The resulting neutrino mass spectrum
\begin{eqnarray}\label{eq:spectrum}
	\{ m, m, 2m \}
\end{eqnarray}
has normal ordering and two degenerate values. In the following we assume small violation of this degeneracy, see Ref.~\cite{Zhuridov:2014vfa} for more details.

We should stress that the mixing matrix in Eq.~\eqref{eq:U} is naturally formed by the eigenvectors of $M$ (the last column of $U$ is the eigenvector, which corresponds to the heavier neutrino mass state), and is different from the popular tri-bimaximal and ``democratic" mixing patterns (see Ref.~\cite{Garg:2013xwa} and references therein), which fail to explain the solar neutrino data by the incoherence, as we do in section~\ref{Solar neutrinos: decoherence}.

%%%%%%%%%%%%%%%%%%%%%%%%%%%%%%%%%%%%%%%%%%%%%%%
\section{Neutrino oscillations}\label{sec:oscillations}

The momentum of a massive neutrino $\nu_i$, which has the mass $m_i$, can be expanded in powers of $(m_i/E)^2$ up to the first order as
\begin{eqnarray}%\label{eq:L_osc}
 		\vec p_i	\approx	\vec p +  m_i^2 \left.\frac{\partial \vec p_i}{\partial m_i^2}\right|_{m_i=0}	=	\vec p - \vec\xi \, \, \frac{m_i^2}{2E},
\end{eqnarray}
where $\vec \xi$ is the vector, which depends on the production process, $\vec p$ and $E=|\vec p|$ are the momentum and energy of a neutrino in the massless approximation, respectively. 
When the oscillation phase is measurable, i.e., $\Delta m_{ij}^2L/(2E)\sim1$, the oscillation probability for the neutrino travelled space-time interval $(\vec L,T)$ at the lowest order in $m_i^2/E^2$ can be written as~\cite{Zralek:1998rp,Giunti:2007ry} %\cite{Lipari}
\begin{eqnarray}\label{eq:Posc}
 &&P_{\nu_\alpha\to\nu_\beta}(L,E) 	= 	\sum_{i,j} U_{\alpha i}^*U_{\beta i}U_{\alpha j}U_{\beta j}^* 
\exp\left( -i2\pi\frac{L}{L^\text{osc}_{ij}} \right)	  E_\text{coh} \,	E_\text{loc}	\nonumber\\
 	&&				=	\sum_i|U_{\alpha i}|^2|U_{\beta i}|^2 + 2\sum_{i>j} \text{Re}\left[U_{\alpha i}^*U_{\beta i}U_{\alpha j}U_{\beta j}^* 
\exp\left( -i2\pi\frac{L}{L^\text{osc}_{ij}} \right)	\right]  E_\text{coh} \,	E_\text{loc}
\end{eqnarray}
with the coherence term
\begin{eqnarray}
 	 E_\text{coh} =	\exp\left[ -\left(\frac{L}{L^\text{coh}_{ij}}\right)^2 \right] 
\end{eqnarray}
and the localization term
\begin{eqnarray}
 	 E_\text{loc} =	\exp\left[ -2\pi^2 \left( 1- \frac{\vec L \cdot\vec\xi}{L} \right)^2\left(\frac{\sigma_x}{L^\text{osc}_{ij}}\right)^2 \right],
\end{eqnarray}
where  
\begin{eqnarray}\label{eq:L_osc}
 L^\text{osc}_{ij}=\frac{4\pi E}{\Delta m_{ij}^2}
\end{eqnarray}
and \begin{eqnarray}\label{eq:L_coh}
 L^\text{coh}_{ij}=\frac{4\sqrt{2} E^2}{|\Delta m_{ij}^2|} \sigma_x
\end{eqnarray}
are the neutrino oscillation and coherence lengths, respectively; $\Delta m_{ij}^2=m_i^2-m_j^2$ is the neutrino mass splitting, and $\sigma_x$ is the neutrino wave packet spatial size.

%%%%%%%%%%%%%%%%%%%%%%%%%%%%%%%%%%%%%%%%%%%%%%%
\subsection{Atmospheric neutrinos:  oscillations} \label{Atmospheric neutrinos:  oscillations}

In the limits 
\begin{eqnarray}\label{eq:limit1}
	L \ll L^\text{coh}_{ij}
\end{eqnarray}
and
\begin{eqnarray}\label{eq:limit2}
	\sigma_x\ll L^\text{osc}_{ij}
\end{eqnarray}
Eq.~\eqref{eq:Posc} can be rewritten as
\begin{eqnarray}
	P_{\nu_\alpha\to\nu_\beta}(L,E)	=	\sum_i  |U_{\alpha i}|^2 |U_{\beta i}|^2	+	2\sum_{i>j}  |U_{\alpha i}^*U_{\beta i} U_{\alpha j}U_{\beta j}^*|	\cos\left(\frac{\Delta m_{ij}^2}{2E}L-\phi_{\beta\alpha;ij}\right)
\end{eqnarray}
with $\phi_{\beta\alpha;ij}	=	\arg(U_{\alpha i}^*U_{\beta i} U_{\alpha j}U_{\beta j}^*)$.  

For the neutrino mass spectrum in Eq.~\eqref{eq:spectrum} and mixing in Eqs.~\eqref{eq:U} and \eqref{eq:Uparametr} we have
\begin{eqnarray}\label{eq:PoscModel}
	P_{\nu_e\to\nu_\tau}(L,E)	=	P_{\nu_\mu\to\nu_\tau}(L,E)	&=&	4 s_{12}^2 c_{23}^2	s_{23}^2	\sin^2 \left( 	\frac{\Delta m^2 L}{ 4E}	\right)
							=	\frac{4}{9}		\sin^2 \left( 	\frac{\Delta m^2 L}{ 4E}	\right),  \\
	P_{\nu_e\to\nu_\mu}(L,E)	&=&	4 c_{12}^2 s_{12}^2	s_{23}^4	\sin^2 \left( 	\frac{\Delta m^2 L}{ 4E}	\right)
							=	\frac{4}{9}		\sin^2 \left( 	\frac{\Delta m^2 L}{ 4E}	\right), 	\label{eq:PoscModel2}
\end{eqnarray}
where $\Delta m^2\equiv m_3^2-m_{i<3}^2 =3m^2$. The oscillations among all the three flavor neutrinos have same size in vacuum. Using the atmospheric neutrino mass splitting $\Delta m_a^2=(2.06-2.67)\times10^{-3}$~eV$^2$ (at 99.73\% CL)% with the best-fit value $\Delta m_a^2=2.35\times10^{-3}$~eV$^2$
~\cite{PDG2012}~\footnote{We consider this result as approximate since it was derived in a very different mixing framework.}, we have ${0.026~\text{eV}<m<0.030~\text{eV}}$.

The flux of atmospheric neutrinos was measured by the Super-Kamiokande (SK) detector~\cite{Ashie:2005ik,Ashie:2004mr,Abe:2006fu} with high statistics for the neutrino energies in the range from 0.1 GeV to $10^3$~GeV. 
Using Eq.~\eqref{eq:L_osc}, we have $L_{32}^\text{osc}(E>0.1~\text{GeV})>100$~km.
Using the estimate $\sigma_x \lesssim m_\mu c\tau_\mu / (4E_\nu)$~\cite{Farzan:2008eg}, where $m_\mu=105.6$~MeV and $c\tau_\mu=658.6$~m are the mass and mean free path of muon, respectively, we have $\sigma_x (E>0.1~\text{GeV}) < 200$~m. 
Hence Eq.~\eqref{eq:limit2} is perfectly satisfied.  Using Eq.~\eqref{eq:L_coh}, one can see that Eq.~\eqref{eq:limit1} is also well satisfied.

We notice that the difference between the $e$-like and $\mu$-like event distributions in the SK experiment~\cite{Ashie:2005ik} can be explained by the matter effect on $\nu_e$ which travel through the Earth. Indeed, 
$P_{\nu_e\to\nu_\tau}$ in Eq.~\eqref{eq:PoscModel} and $P_{\nu_e\to\nu_\mu}$ in Eq.~\eqref{eq:PoscModel2} can be uniformly written as
%considered as the transition probability in the  2-neutrino mixing scheme 
\begin{eqnarray}
	P(\nu_e\to\nu_x)	=	\sin^22\theta  \sin^2 \left( 	\frac{\Delta m^2 L}{ 4E}	\right)
\end{eqnarray}
with $x=\mu,\tau$, and  $\sin2\theta = 2/3$, where $\theta$ is the neutrino mixing angle in vacuum. For the neutrinos propagating in the matter with the electron number density $N_e$ this probability will be modified as~\cite{PDG2012,Wolfenstein:1977ue,Mikheev:1986gs,Barger:1980tf}
\begin{eqnarray}
	P_m(\nu_e\to\nu_x)	=	\sin^22\theta_m  \sin^2 \left( 	\frac{\Delta M^2 L}{ 4E}	\right)
\end{eqnarray}
with $\Delta M^2 = \Delta m^2 F_m$ and
\begin{eqnarray}\label{eq:sin2matt}
	\sin2\theta_m  = 	\frac{\tan2\theta}{\sqrt{\left(1 - \frac{N_e}{N_e^{\rm res}} \right)^2 + \tan^22\theta }},
\end{eqnarray}
where the matter factor is given by
\begin{eqnarray}%\label{eq:sin2matt}
	F_m		=	\left[	\left(1 - \frac{N_e}{N_e^{\rm res}} \right)^2 \cos^22\theta + \sin^22\theta	\right]^{\frac{1}{2}},
\end{eqnarray}
and the Mikheyev-Smirnov-Wolfenstein ``resonance density" is given by
\begin{eqnarray}\label{eq:res_density}
	N_e^{\rm res}	=	\frac{\Delta m^2 \cos2\theta}{2\sqrt{2}E G_\text{F}}  \approx 	6.56\times10^6 \,	\frac{\Delta m^2 [\text{eV}^2]}{E [\text{MeV}]}	\cos2\theta	\quad	\text{cm}^{-3} \, \text{N}_\text{A}
\end{eqnarray}
with $G_\text{F}$ and $\text{N}_\text{A}$ being Fermi constant and Avogadro number, respectively. 
In particular, for the atmospheric neutrinos with energies $E\sim10$~GeV we have $N_e^{\rm res} \approx 1.15~\text{cm}^{-3} \, \text{N}_\text{A}$, and using the mean electron number density in the Earth core $\bar N_e^c \approx 5.4~\text{cm}^{-3} \, \text{N}_\text{A}$~\cite{PDG2012,Dziewonski:1981xy}, we find the oscillation probabilities
\begin{eqnarray}
	P_m(\nu_e\to\nu_x)	=	0.05 \,  \sin^2 \left( 	2.8\frac{\Delta m^2 L}{ 4E}	\right),
\end{eqnarray}
which are significantly suppressed with respect to $P_{\nu_\mu\to\nu_\tau}$ in Eq.~\eqref{eq:PoscModel}. The corresponding suppression in the Earth mantle is less significant. More detailed investigation of the Earth's matter effect on the atmospheric neutrinos in the democratic model (with possible small perturbations) is in progress~\cite{DZ}.

%%%%%%%%%%
\begin{figure}
  \centering
  \includegraphics[width=.6\textwidth]{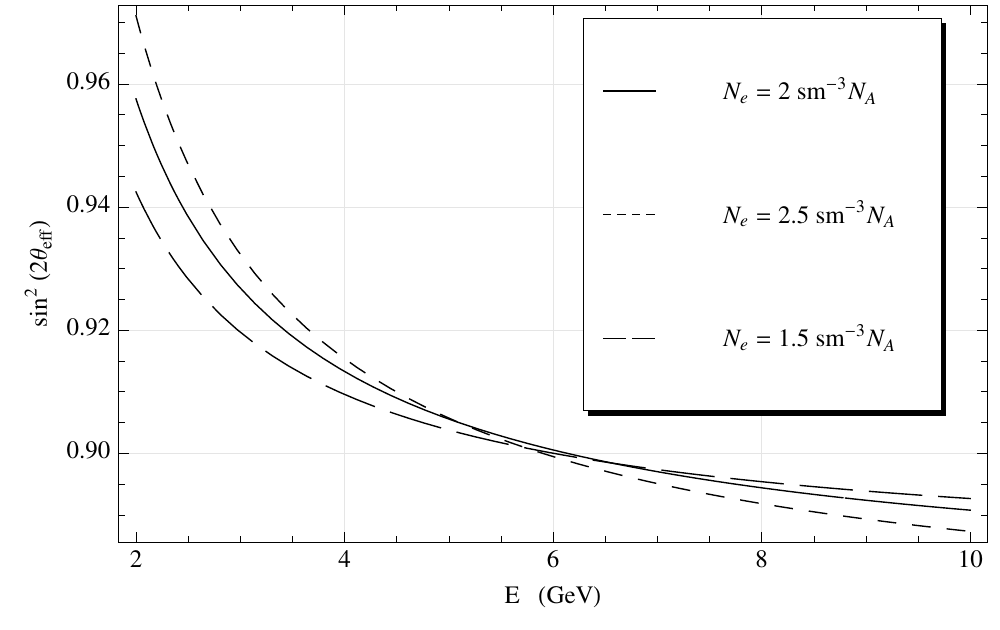}
   \caption{Effective factor $\sin^22\theta_\text{eff}$ in the muon neutrino survival probability versus the neutrino energy for the MINOS baseline $L=734$~km and chosen values of the electron number density $N_e$.}\label{Fig:MINOS}
\end{figure}
The MINOS result for the amplitude of the muon neutrino oscillations~\cite{Adamson:2013whj,Adamson:2011ig}, which is close to unity, is explained by the muon neutrino oscillations to both electron neutrino and tau neutrino, where the matter effect plays subdominant role. 
For the survival probability of the muon neutrino we have
\begin{eqnarray}\label{eq:PoscSurvival}
	P^\text{surv}(\nu_\mu\to\nu_\mu)	&=&	1	-	P(\nu_\mu\to\nu_\tau)	-	P_m(\nu_\mu\to\nu_e)	\nonumber\\
								&=&	1	-	\frac{4}{9}	\sin^2 \left( \phi_\text{osc}(L,E)\right)		-	\sin^22\theta_m \sin^2 \left(\phi_\text{osc}(L,E) F_m \right)
\end{eqnarray}
with the vacuum oscillation phase
\begin{eqnarray}%\label{eq:PoscModel}
	 \phi_\text{osc}(L,E)	\equiv	\frac{\Delta m^2 L}{ 4E} 	=	1.27\frac{\Delta m^2[\text{eV}^2] L[\text{km}]}{ E[\text{GeV}]}.
\end{eqnarray}
Eq.~\eqref{eq:PoscSurvival} can be effectively rewritten as
\begin{eqnarray}%\label{eq:PoscSurvival}
	P^\text{surv}(\nu_\mu\to\nu_\mu)	=	1	-	\sin^22\theta_\text{eff}(L,E,N_e) \sin^2 \left(\phi_\text{osc}(L,E) \right)
\end{eqnarray}
with
\begin{eqnarray}%\label{eq:PoscSurvival}
	\sin^22\theta_\text{eff}(L,E,N_e) 	=	\frac{4}{9}		+	\sin^22\theta_m(E,N_e) \, 	\frac {\sin^2 \left(\phi_\text{osc}(L,E) F_m(N_e) \right)}{\sin^2 \left(\phi_\text{osc}(L,E) \right)}.
\end{eqnarray}
Fig.~\ref{Fig:MINOS} shows the dependence of $\sin^22\theta_\text{eff}$ on the muon neutrino energy $E$ in the range from 2~GeV to 10~GeV  for the MINOS baseline $L=734$~km, where the differences among the three lines demonstrate the relatively weak dependence on $N_e$ (in the Earth crust).

%%%%%%%%%%%%%%%%%%%%%%%%%%%%%%%%%%%%%%%%%%%%%%%%%%%%%%%%%%%%%
\subsection{Solar neutrinos: decoherence} \label{Solar neutrinos: decoherence}

Thousands of $^8B$ solar neutrinos were detected by Sudbury Neutrino Observatory (SNO)~\cite{Aharmim:2005gt} using the charged-current (CC) and neutral current (NC) reactions
\begin{eqnarray}
	\nu_e+d\to	e^-+p+p
\end{eqnarray}
and
\begin{eqnarray}
	\nu_\ell+d\to	\nu_\ell+p+n,
\end{eqnarray}
respectively. 
The ratio of the neutrino fluxes measured with  CC and NC events is~\cite{Aharmim:2005gt,PDG2012}
\begin{eqnarray}\label{eq:CC/NC}
 \frac{\Phi_\text{SNO}^\text{CC}}{\Phi_\text{SNO}^\text{NC}} = \frac{1.68\pm0.06^{+0.08}_{-0.09}}{4.94\pm0.21^{+0.38}_{-0.34}} = 0.340^{+0.074}_{-0.063}.
\end{eqnarray}

For the solar neutrinos with the energies $E\lesssim10$~MeV the oscillations due to $\Delta m_a^2$ proceed in the matter of the Sun as in vacuum, since the number density of electrons in the center of the Sun is by a factor of 10 smaller than the ``resonance density" in Eq.~\eqref{eq:res_density} \cite{PDG2012}.
The wave packet size of solar neutrinos, which have the energies $E\sim1$~MeV, is $\sigma_x\sim10^{-9}$~m~\cite{Kiers:1995zj}.  Using Eq.~\eqref{eq:L_coh} with $\Delta m_a^2$, we have $L^\text{coh}_{31} \sim 10^3$~km. For the Earth-Sun distance $L\approx 1.5\times10^8$~km we have 
\begin{eqnarray}
	L^\text{coh}_{31} \ll L.
\end{eqnarray}
(This relation is satisfied for the neutrino mass splittings down to $\sim10^{-8}$~eV$^2$, which sets the lower bound for the violation of the degeneracy in the neutrino mass spectrum in Eq.~\eqref{eq:spectrum}  without spoiling the result in Eq.~\eqref{eq:solar_nu_flux_suppression}.) 
In this limit Eq.~\eqref{eq:Posc} is reduced to the simple incoherent form
\begin{eqnarray}\label{eq:incoh}
	P_{\nu_\alpha\to\nu_\beta}^\text{incoh}	=	\sum_i  |U_{\alpha i}|^2 |U_{\beta i}|^2.
\end{eqnarray}
(Moreover the oscillations should be averaged out already due to the lack of the emitter localization~\cite{Gribov:1968kq,Nussinov:1976uw}.) 
Hence 
\begin{eqnarray}
 \frac{\Phi_\text{sol}^\text{CC}}{\Phi_\text{sol}^\text{NC}} =		\frac{P_{\nu_e\to\nu_e}^\text{incoh}}{\sum_\beta P_{\nu_e\to\nu_\beta}^\text{incoh}}	=	 \sum_i|U_{ei}|^4,
\end{eqnarray}
and, using Eq.~\eqref{eq:U}, we have
\begin{eqnarray}\label{eq:solar_nu_flux_suppression}
	\sum_i|U_{ei}|^4 =	\frac{7}{18}	\simeq	0.39,
\end{eqnarray}
which is in good agreement with the experimental data in Eq.~\eqref{eq:CC/NC}.

We remark that the suppression of $\nu_e$ flux in Eq.~\eqref{eq:solar_nu_flux_suppression} is independent of their energy. This provides an efficient tool for examination of the solar model calculations, since the neutrino fluxes are quite sensitive to their parameter variations. 
For example, the predictions of the solar models with late accretion for $^8$B, $^7$Be and $pp$ neutrino fluxes differ by 400\%, 300\% and 20\%, respectively, see Fig.~12 in Ref.~\cite{Serenelli:2011py}.

We notice that the explanation of KamLAND experimental data by the neutrino oscillations with the mass splitting $\Delta m_\nu^2\sim10^{-5}$~eV$^2$~\cite{Araki:2004mb,Gando:2010aa,Gando:2013nba}  is questionable\footnote{Superposition of the neutrino fluxes from various reactors of different power can, in principle, explain the shape of the KamLAND signal and the difference between pre-Fukushima~\cite{Gando:2010aa} and post-Fukushima~\cite{Gando:2013nba} results without the use of the neutrino oscillations with $\Delta m_\nu^2\sim 10^{-5}$~eV$^2$.}, and at least one independent experiment is needed to verify the KamLAND result.

%%%%%%%%%%%%%%%%%%%%%%%%%%%%%%%%%%%%%%%%%%%%%%%
\section{Other neutrino experiments}\label{sec:other_experiments}

%%%%%%%%%%%%%%%%%%%%%%%%%%%%%%%%%%%%%%%%%%%%%%%
\subsection{Direct neutrino mass experiments} % oscillations + localization?

Using Eqs.~\eqref{eq:U} and \eqref{eq:spectrum}, the average mass, determined through the analysis of low energy beta decays, can be written as
\begin{eqnarray}
	\langle m_\beta \rangle 	\equiv	\sqrt{ \sum_i m_i^2 |U_{ei}|^2 }	 =	m\sqrt{2}.
\end{eqnarray}
For $m\simeq0.03$~eV we have $\langle m_\beta \rangle \simeq0.04$~eV, which is far below the present upper limit $\langle m_\beta \rangle < 2.5$~eV (at 95\% CL)  obtained in the ``Troitsk Neutrino Mass" Experiment~\cite{Lobashev:2001uu}, and below the sensitivity $0.2$~eV (at 90\% CL) of ongoing KATRIN experiment~\cite{Titov:2004pk,Formaggio:2012zz}.  However the new approaches such as the MARE, ECHO, and Project8 experiments, offer the promise to perform an independent measurement of the neutrino mass in the sub-eV region~\cite{Drexlin:2013lha}.

%%%%%%%%%%%%%%%%%%%%%%%%%%%%%%%%%%%%%%%%%%%%%%%
\subsection{Neutrinoless double beta decay (\boldmath{$0\nu2\beta$})} 

Using Eqs.~\eqref{eq:U} and \eqref{eq:Lmass}, the effective Majorana mass in $0\nu2\beta$ decay can be written as
\begin{eqnarray}
	\langle m\rangle 	\equiv	\sum_i s_i m_iU_{ei}^2	=	0.		
\end{eqnarray}
Hence $0\nu2\beta$ decay is strongly suppressed and practically unobservable in the discussed model, which can be checked to very good precision in possible future ton-scale experiments, e.g., GENIUS~\cite{Hellmig:1998pv}, LXe 10ty and HPXe 10ty~\cite{GomezCadenas:2013ue}.

%%%%%%%%%%%%%%%%%%%%%%%%%%%%%%%%%%%%%%%%%%%%%%%
\subsection{\boldmath{$\nu_\mu\leftrightarrow\nu_e$} oscillations}

The T2K experiment reported indication of electron neutrino appearance from an accelerator-produced off-axis muon neutrino beam~\cite{Xing:2011at,Abe:2011sj}.  
The MINOS Collaboration also reported $\nu_\mu\to\nu_e$ appearance signal~\cite{Adamson:2011qu}.  
The discussed T2K (MINOS) experiment observed 6 (62) candidate $\nu_e$ events against the expected 1.5 (50) events for the case of no  $\nu_\mu\to\nu_e$ oscillation, which resulted from their analysis, in particular, in the upper bounds $A \lesssim 0.14~(A \lesssim 0.06)$ at the 90\% CL on the amplitude of the oscillation probability written as
\begin{eqnarray}
	P(\nu_\mu\to\nu_e)	\approx	A  \sin^2 \left( 	\phi_\text{osc}(L,E)	\right).
\end{eqnarray}
More data is expected from ongoing and coming experiments such as NOvA~\cite{Ayres:2004js}.
%However the data collected in 2008 and 2009 in the OPERA experiment are compatible with the non-oscillation hypothesis~\cite{Agafonova:2013xsk}.  

Several reactor experiments recently reported indications of electron antineutrino disappearance~\cite{Abe:2011fz,An:2012eh,Ahn:2012nd} with the ratios of observed to expected numbers of $\bar\nu_e$ about $R=0.92-0.94$, which yielded $A\simeq0.05$. 

The discussed values of the amplitude $A$ are smaller than the predictions, which can be derived from the theory of democratic neutrino oscillations, introduced in section~\ref{Atmospheric neutrinos:  oscillations}.  
However these results are strongly model dependent. In particular, 
%the results of the accelerator neutrino experiments depend on the assumption of $\sin^22\theta_{23}=1$. 
%Use of $\sin^22\theta_{23}<1$ should increase the resulting values of $A$. 
%Also 
the uncertainties  of the neutrino fluxes may have a significant effect, e.g., reduction of $R$ by about 10\% results in increase of $A$ by factor of 2.5. 
Therefore we strongly encourage the neutrino physicists to reanalyze the data in the framework of the presented simple and natural theory. 
We hope that this will bring the neutrino theory and experiment to a good agreement without introduction of sterile neutrinos or other exotics.

%%%%%%%%%%%%%%%%%%%%%%%%%%%%%%%%%%%%%%%%%%%%%%%
\section{Conclusion} 

We introduced a simple model, which may explain the confirmed neutrino data, and resolve the long standing neutrino mass puzzles. This model is based on the non-diagonal ``democratic" mass matrix of three Majorana neutrinos, and predicts degenerate neutrino mass splittings, which govern the atmospheric neutrino oscillations, and result in the neutrino mass spectrum close to \{0.03, 0.03, 0.06\}(eV). The difference between $e$-like and $\mu$-like event distributions in the Super-Kamiokande is controlled by the matter effect. The solar neutrino deficit is explained by their incoherence in the Earth. We outlined the predictions for the direct neutrino mass and the neutrinoless double beta decay experiments, and importance of the accelerator $\nu_\mu\to\nu_e$ oscillation and the reactor $\bar\nu_e$ disappearance experiments in verification of the considered model. Definitely more detailed and careful analysis of the neutrino data with respect to the considered model will be useful.

%%%%%%%%%%%%%%%%%%%%%%%%%%%%%%%%%%%%%%%%%%%%%%%
\section*{Acknowledgements} 

This work was supported in part by the US Department of Energy under the contract DE-SC0007983. The author  
thanks Lincoln Wolfenstein, Anatoly Borisov and the PRD Referee for useful comments, and Alexey Petrov and Gil Paz for important questions.

\end{document}